\documentclass[conference]{IEEEtran}

\usepackage{balance}
\usepackage{booktabs}

\usepackage{graphicx}
\DeclareGraphicsExtensions{.pdf,.jpeg,.png}

\usepackage[hidelinks=true]{hyperref}
\usepackage{tabularx}
\usepackage{xcolor}
\usepackage[binary-units]{siunitx}
\usepackage{setspace}
\usepackage{acronym}
\usepackage{rotating}
\usepackage[tight]{subfigure}
\usepackage{units}
\usepackage{multirow}
\usepackage{listings}

\usepackage[inline]{enumitem}

\newcommand{\comment}[1]{}

\newacro{VM}{virtual machine}
\newacro{MPU}{memory protection unit}

\newcommand\blfootnote[1]{%
  \begingroup
  \renewcommand\thefootnote{}\footnote{#1}%
  \addtocounter{footnote}{-1}%
  \endgroup
}

\begin{document}

\title{
\footnotesize
\framebox[1.01\width]{\parbox{\dimexpr\linewidth-2\fboxsep-2\fboxrule}{If you cite this paper, please use the PEMWN 2020 reference: K. Zandberg, E. Baccelli. Minimal Virtual Machines on IoT Microcontrollers: The Case of Berkeley Packet Filters with rBPF. In Proc. of 9th IFIP/IEEE PEMWN, Dec. 2020.}}
 \ \\ \ \\ \ \\
\Huge Minimal Virtual Machines on IoT Microcontrollers: \\ The Case of Berkeley Packet Filters with rBPF}

%

\author{
\IEEEauthorblockN{Koen Zandberg\IEEEauthorrefmark{1},
Emmanuel Baccelli\IEEEauthorrefmark{1}\IEEEauthorrefmark{2}
}
\\\
    \IEEEauthorblockA{\IEEEauthorrefmark{1}Inria, France
}
\IEEEauthorblockA{\IEEEauthorrefmark{2}Freie Universit\"at Berlin, Germany
}
}



\maketitle

\begin{abstract}
Virtual machines (VM) are widely used to host and isolate software modules.
However, extremely small memory and low-energy budgets have so far prevented wide use of VMs on typical microcontroller-based IoT devices.
In this paper, we explore the potential of two minimal VM approaches on such low-power hardware.
We design rBPF, a register-based VM based on extended Berkeley Packet Filters (eBPF).
We compare it with a stack-based VM based on WebAssembly (Wasm) adapted for embedded systems.
We implement prototypes of each VM, hosted in the IoT operating system RIOT\@.
We perform measurements on commercial off-the-shelf IoT hardware. 
Unsurprisingly, we observe that both Wasm and rBPF virtual machines yield execution time and memory overhead, compared to not using a VM\@.
We show however that this execution time overhead is tolerable for low-throughput, low-energy IoT devices.
We further show that, while using a VM based on Wasm entails doubling the memory budget
for a simple networked IoT application using a 6LoWPAN/CoAP stack,
using a VM based on rBPF requires only negligible memory overhead (less than 10\% more memory).
rBPF is thus a promising approach to host small software modules, 
isolated from OS software, and updatable on-demand, over low-power networks.\blfootnote{Acknowledgement: H2020 SPARTA partly funded this work.}

\end{abstract}


%
\IEEEpeerreviewmaketitle

\section{Introduction}

The availability of cheap low-power microcontrollers and low-power radios is driving the emergence of the Internet of Things (IoT).
Typical microcontrollers based on architecture such as Arm Cortex-M are combined with various sensors/actuators, and with a radio such as Bluetooth Low-Energy, LoRa or IEEE 802.15.4, on a small Arduino-like hardware module.
This class of embedded hardware~\cite{rfc7228} trades off limiting resources (slow processor, kilobytes of RAM and Flash memory\ldots), for small energy consumption (in the milliwatt range) and a small price tag (a few dollars).


In parallel, however, security concerns~\cite{neshenko2019demystifying} grow with the emergence of IoT.
Cyberphysical chain reactions~\cite{soltan2018waterheaterbotnet}, or extended functionality attacks~\cite{ronen2016extended} expand the traditional attack surface of networked systems.

To mitigate such a variety of attacks, specific security mechanisms are needed at all levels of the system.
For instance, on-going work defines new network protocols and workflows aiming to mitigate network- and some software-based attack vectors~\cite{tschofenig2019cyberphysical}.


Complementary mechanisms focus on isolating critical software processes (e.g.\ access to specific sections of the address space) from the rest of the application software running on-board the microcontroller.
This facilitates establishing a root of trust on the microcontroller, which can bootstrap other security mechanisms.

Concretely, we consider two categories of use-cases:
\begin{enumerate}
  \item Isolating high-level business logic, updatable on-demand remotely over the low-power network. This type of logic is rather long-lived, and has loose (non-real-time) timing requirements.
  \item Isolating debug/monitoring code snippets at low-level, inserted and removed on-demand, remotely, over the network. Comparatively, this type of logic is short-lived and exhibits stricter timing requirements.
\end{enumerate}

One approach is to modify the hardware architecture of microcontrollers, adding specific hardware mechanisms to guarantee such isolations.
Such hardware functionalities facilitate establishing a root of trust on the microcontroller.
Prominent examples of this trend include TrustZone on Arm Cortex-M architectures~\cite{pinto2019trustzone}, Sanctum on RISC-V architectures~\cite{sanctum-riscv}, Sancus2.0 on MSP430 architectures~\cite{noorman2017sancus}.

However, changing hardware is both
(i) more difficult than upgrading software, and
(ii) heavily dependent, by nature, on a specific hardware architecture.
Therefore, a legitimate question which arises is: what \emph{software-only} equivalent can be achieved, to isolate the processes in our use-cases?


\section{Related Work}

Different categories of software-based process isolation techniques have been developed specifically for microcontrollers.
Small \emph{virtual machines} are used to host and isolate processes from other processes running on the microcontroller. 
For example Darjeeling~\cite{darjeeling} is a subset of the Java VM, modified to use a 16 bit architecture, designed for 8- and 16-bit microcontrollers.
Another example is WebAssembly (Wasm~\cite{haas2017wasm}), a \ac{VM} specification with a stack-based architecture, designed for process isolation in Web browsers, which has recently been ported to microcontrollers~\cite{wasm3}. 
Beyond the low-power IoT domain, tiny \acp{VM} are also used in other contexts for a long time. 
For instance JavaCard~\cite{identity2018javacard} uses a small Java \ac{VM} running on smart cards. 
Elsewhere, in the Linux ecosystem, eBPF~\cite{mccanne1993bpf,fleming2017eBPF} enables a small \ac{VM} hosting and isolating debug and inspection code, in the Linux kernel, at run-time.


Another type of approach uses \emph{scripted logic interpreters} to isolate some processes. 
For instance, prior work such as~\cite{riot-js-container} uses a small JavaScript run-time container, hosting (updateable) business logic, interpreted on-board a microcontroller, glued atop a real-time OS (RIOT).

Yet another category of solution uses \emph{OS-level mechanisms for process isolation}.
For instance, Tock~\cite{tockos} is an OS written in the Rust programming language, which offers strong isolation between its kernel and application logic processes.
However, Tock requires hardware providing an \ac{MPU} (only some Cortex-M and RISC-V hardware is supported so far).

The goal we pursue in this paper is to explore in practice solutions which:
\begin{itemize}
\item require minimal memory footprint;
\item do not depend on extra hardware-specific mechanisms to protect memory;
\item offer tolerable code execution speed slump;
\item require small data transfer over-the-air when isolated code is updated;
\end{itemize}
 
For this purpose, we explore approaches based on \aclp{VM}.
More specifically, we consider two architectures of \acp{VM}:
a stack-based \ac{VM} based on WebAssembly, and
a register-based \ac{VM} based on eBPF, as described below.

%
The main contributions of this paper are:
\begin{itemize}
\item we design rBPF, an adaptation of eBPF providing a software-based solution to isolate processes on low-power microcontrollers;
\item we provide the implementation of two open source prototypes of \acp{VM}, using rBPF on one hand, and on the other hand using Wasm, based on the WASM3 interpreter;
\item we evaluate the performance of our rBPF prototype compared to Wasm, via experiments running the \acp{VM} on real microcontrollers; 
\item we show that rBPF offers promising perspectives in terms of smaller memory footprint, we discuss security guarantees and potential next steps.
\end{itemize}

\section{Background}

\subsection{WebAssembly}
WebAssembly (Wasm~\cite{haas2017wasm}) is a virtual instruction set architecture, standardized by the World Wide Web Consortium (W3C), primarily aimed at portable web applications.
The instruction set allows for binaries small in size, to minimize transfer time to the client.
The sandbox provided by implementations offers strong guarantees on memory access.
Both of these properties aim to ensure security while requiring only limited memory footprint on the platform target.

The WebAssembly \ac{VM} uses both a stack and a flat heap for memory storage.
The stack is required by the architecture, and can be configured to any size.
An interface for allocating heap memory is provided by the standard.
Specification mandates memory allocations in chunks of \SI{64}{\kibi\byte} (pages).

\begin{figure}
  \centering
  \includegraphics[width=0.8\linewidth]{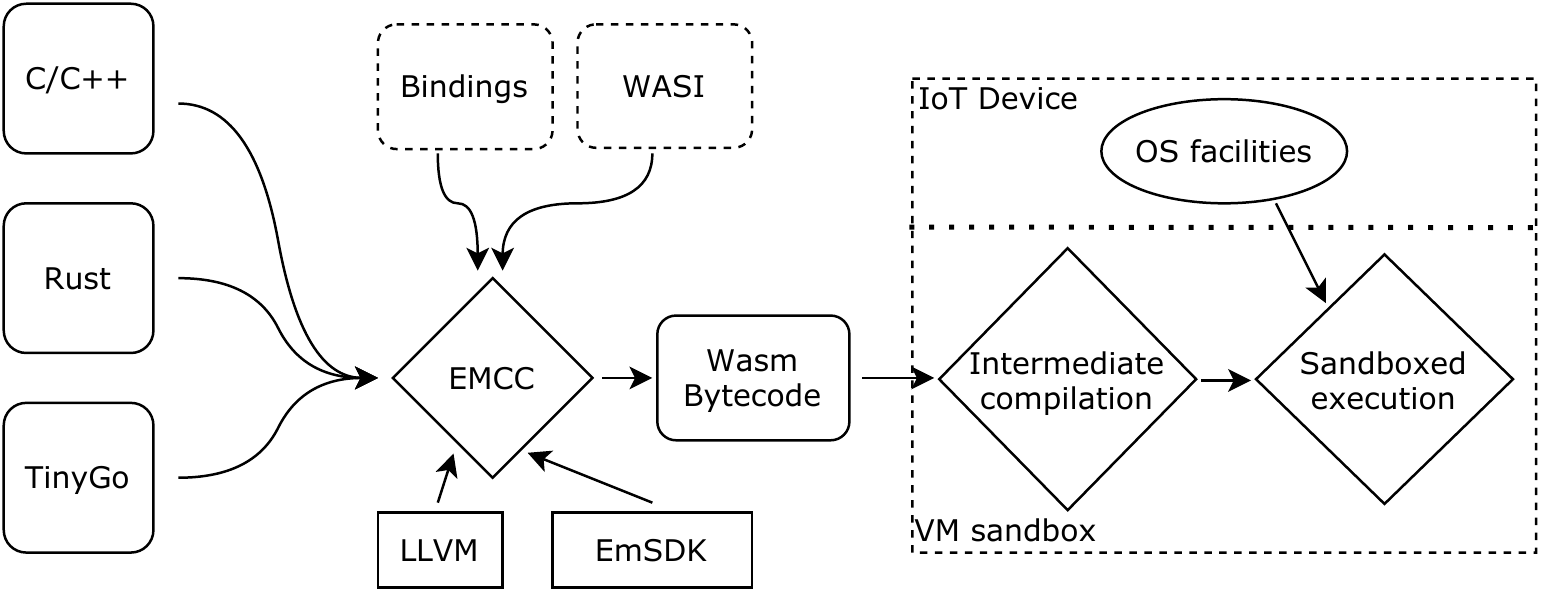}
  \caption{Wasm code development and execution workflow with the WASM3 \ac{VM}.}
  \label{fig:wasm3_workflow}
  \vspace{-1.5em}
\end{figure}

{\bf Toolchain \& SDK}. 
The full workflow for development and execution of Wasm applications is depicted in \autoref{fig:wasm3_workflow}.
Wasm uses the LLVM compiler: applications in any language supported by LLVM are possible, such as C/C++, D, Rust, and TinyGo among others.
A standardized interface is specified for host access in a POSIX-like way is provided by the WASI standard~\cite{WASI}.

{\bf Interpreter.} Once the Wasm binary is created, it can be transferred to the IoT device,
on which it is interpreted and executed, as shown in \autoref{fig:wasm3_workflow}.
Several interpreters exist,
in this paper we use the WASM3~\cite{wasm3} interpreter,
which uses a two-stage approach: the loaded application is first transpiled to an optimized executable, which then can be executed in the interpreter.

\subsection{Extended Berkeley Packet Filters}

Berkeley Packet Filter (BPF~\cite{fleming2017eBPF}) is a small in-kernel \ac{VM} available on most Unix-like operating systems.
Its original purpose was network packet filtering, for example: only pass to userspace packets matching a set of requirements.
Within the Linux kernel the \ac{VM} is extended (to eBPF) to allow for multiple non-network related purposes.
eBPF provides a small and efficient facility for running custom code inside the kernel, hooking into various subsystems.

The state-of-the-art eBPF architecture is 64-bit register based \ac{VM} with a fixed stack.
The stack itself is specified as fixed at \SI{512}{\byte}.
A heap is not contained in the specification. As an alternative, the Linux kernel provides an interface to key-value maps for persistent storage between invocations.
The limited stack size and absence of a heap put only minimal requirements on the RAM a platform has to provide for the \ac{VM}.

\begin{figure}
\centering
\includegraphics[width=0.8\linewidth]{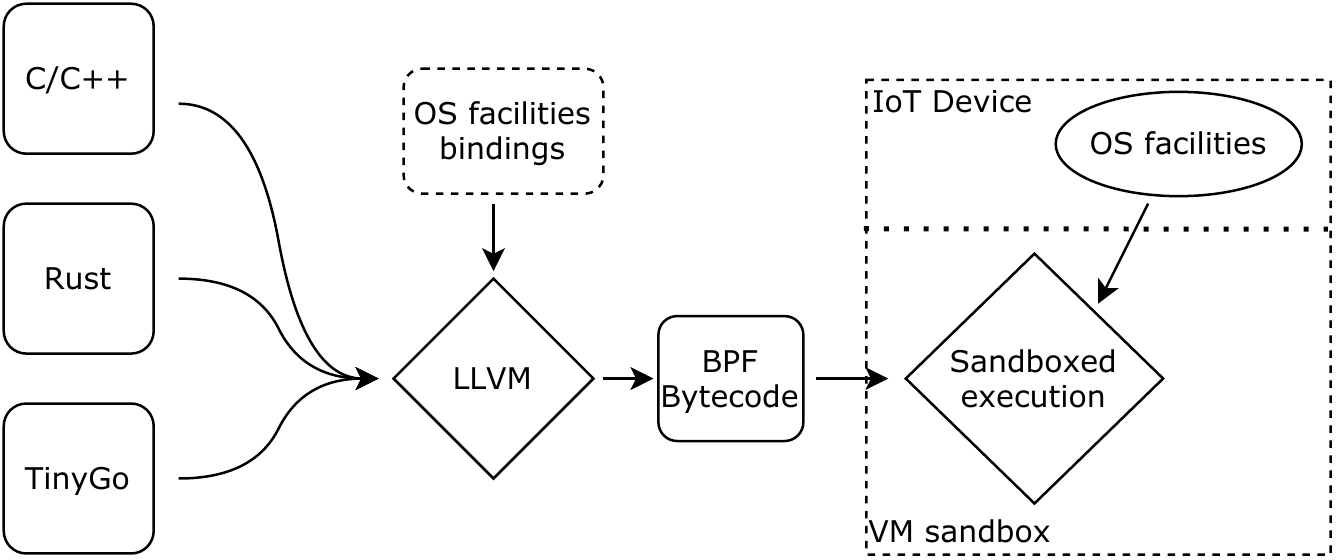}
\caption{rBPF code development and execution workflow.}
\label{fig:rbpf_workflow}
\vspace{-1.5em}
\end{figure}

The \ac{VM} itself is inherently suitable for isolating the operating system from the virtualized application:
all memory access, including to the stack, happen via load and store instructions.
Moreover, branch and jump instructions are also limited, the application has no access to the program counter and a jump is always direct and relative to current program counter.
The \ac{VM} does not provide facilities to directly write the program counter.
Both of these potential attack surfaces can be implemented with the necessary checks in place to limit access and execution.

Interfacing with the operating system facilities can be done by providing the necessary bindings on the device.

{\bf Comparing eBPF to WebAssembly.}
As with WebAssembly, eBPF makes use of the LLVM toolchain for compilation (see \autoref{fig:rbpf_workflow}) thus any language supported by LLVM can be used and compiled to bytecode.
However, there are differences in terms of architecture, and in terms of memory model.
The architecture of WebAssembly reduces the size of instructions significantly.
On the other hand, eBPF instructions are always \SI{64}{\bit} in size, filled with zero bits where a field is not used.
The stack-based with heap memory approach from WebAssembly put significant requirements on the available RAM\@.
On the other hand, with eBPF, the few registers combined with the limited stack put only minimal pressure on the available RAM\@.

\section{rBPF Design}
The rBPF \ac{VM} is a variant of the eBPF \ac{VM}, designed to be ISA compatible with eBPF\@.
The main difference between rBPF and eBPF lies in the bindings provided for access to the operating system facilities and the events by which execution is triggered. In this paper, we chose RIOT~\cite{baccelli2018riot} as OS to host our VM prototype, but this choice is arbitrary:  our approach could apply to another OS in the same category.

\begin{figure}[!t]
  \centering
  \includegraphics[width=0.7\linewidth]{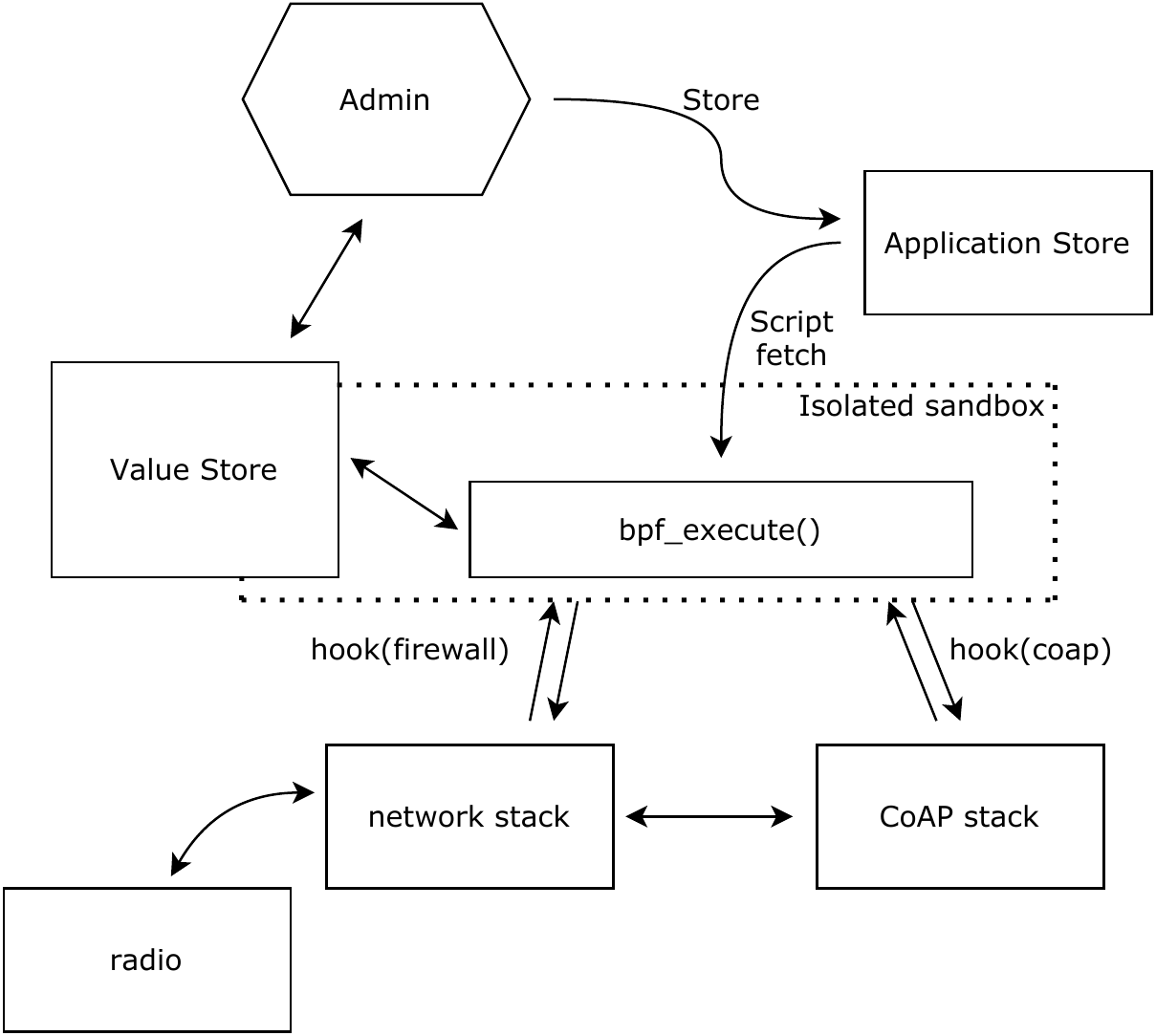}
  \caption{VM integration \& sandboxed execution of rBPF in host OS.}
\label{fig:integration}
  \vspace{-1.75em}
\end{figure}

{\bf VM integration in the OS.}
The rBPF virtual machine is integrated in RIOT as shown in \autoref{fig:integration}.
Within the operating system the \ac{VM} is scheduled as a regular thread, restricted by the scheduler to the configured run priority.
The \ac{VM} does not interfere with real-time thread execution on the operating system.
However, running real-time constrained applications inside the \ac{VM} is not suitable.

As shown in \autoref{fig:integration}, multiple OS event sources can trigger the execution of an application.
For example a request received on the CoAP server or packets passing through the network stack.
Each of these event types can trigger a different rBPF application from the application store, configured by the device administrator.
Similar to eBPF the \ac{VM} supports both an argument passed to the application and a return code from the application back to the calling event.
This can be used to communicate vital execution context with the application and pass a return value back to the initiator.
Further integration with the operating system is available through function bindings, including access to facilities relevant to IoT applications such as sensor values and network packet creation.
With these capabilities the \ac{VM} application, while isolated from the operating system, it retains enough flexibility to host business logic application or simple measure and debug applications.

The application running inside the \ac{VM} is expected to be short-lived, updating an intermediate result or formatting a response to a request.
To provide persistent data between these short-lived invocations a key-value store is available.
An application can read and write values to both a global and a per-script local storage.
Counters or aggregate sensor values can be stored for retrieval in a subsequent execution or queried from outside the \ac{VM} by the operating system.



\begin{figure}[!t]
  \centering
  \includegraphics[width=0.7\linewidth]{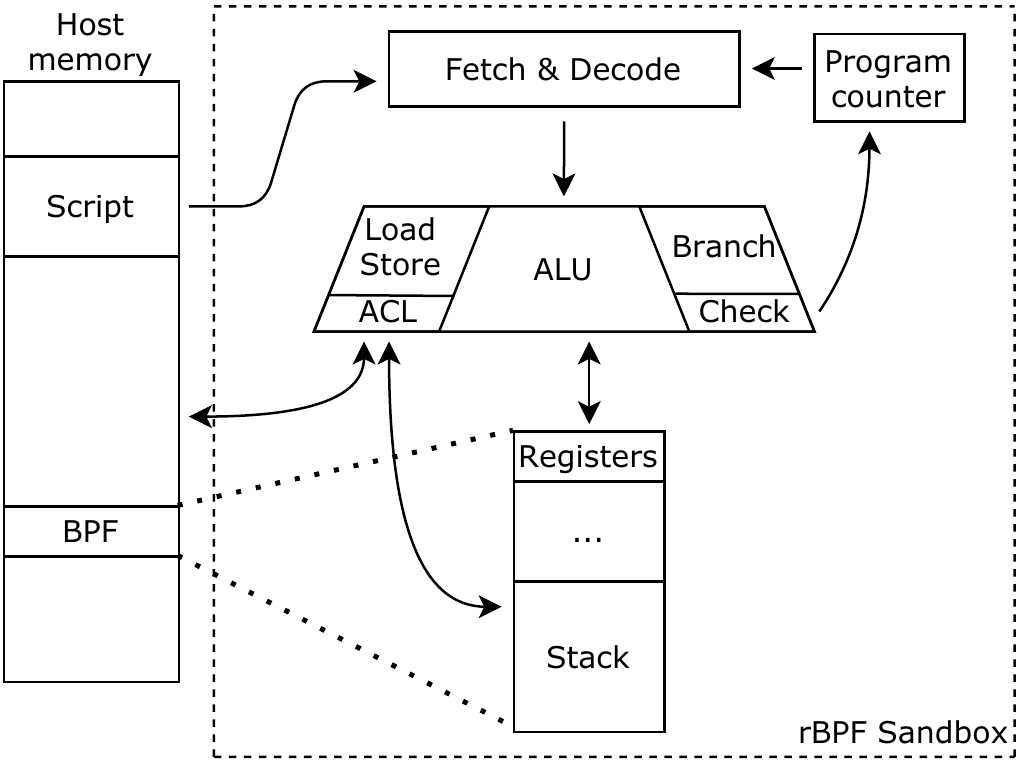}
  \caption{rBPF execution and memory architecture}
  \label{fig:architecture}
  \vspace{-1.25em}
\end{figure}

{\bf VM execution sandboxing.}
The \ac{VM} is based on an iterative loop design, iterating over the application instructions depicted in \autoref{fig:architecture}, which shows
the interaction between the instructions, sandbox guards, and the host address space.
Both the registers and the application stack reside in the memory of the host.
Depending on the instruction to be executed, different protection mechanisms are activated.
Two main protection mechanism are in place to isolate the code executed in the \ac{VM}.

First, the host address space is isolated from the sandbox by access policies loaded in the \ac{VM}.
Every memory access, including stack reads and writes, are subjected these access policy rules.
Different address space sections can be configured to allow reads, writes or both by the caller of the \ac{VM}.
This offers minimal overhead for memory access while providing the guarantees required for the sandbox.

Second, protections on the code executed ensure the \ac{VM} does not start executing code outside the supplied application, such as gadgets deployed by an attacker.
The mechanism works by guarding the branch and jump instructions, ensuring that the destination is not outside the application address space.
As the eBPF ISA implemented does not support indirect jump instructions and no program counter register is available,
the only mechanism to modify the virtual program counter is via the already guarded direct branch and jump instructions.
While the lack of an indirect jump instruction does somewhat limit the flexibility of the applications,
it ensures that all jump destinations are known before executing the application, making a preflight validation of all jump instructions a viable option.







\section{Experimental setup}

We carry out our measurements on popular, commercial off-the-shelf IoT hardware: a Nordic nRF52840 Development Kit, which provides a typical microcontroller (Arm Cortex-M4) with 256 KB RAM, 1 MB Flash, and a 2.4 GHz radio transceiver compatible both with IEEE 802.15.4, and Bluetooth Low-Energy.
This hardware is also available for reproduceability on open access testbeds such as IoT-Lab~\cite{iot-lab2015-ieee-iot-wf}.

On this platform, we perform two types of benchmarks.
First, we perform embedded computing hosted in the VM, to get an idea of basic VM performance.
Then, we perform further benchmarks involving IoT networking capabilities used from within the \ac{VM}.

\subsection{Computing Benchmark Setup}

First, we benchmark a setup consisting of a Fletcher32 checksum algorithm~\cite{fletcher}.
The Fletcher32 checksum algorithm requires a mix of mathematical operations memory reads and branches, containing a loop over input data.
Benchmark results consist of the impact of the \ac{VM} on the operating system in the additional memory required to include it.
For the \acp{VM} themselves, the execution speed and the size of compiled applications loaded into the \ac{VM} is measured.

\if{0}
binary statistics: 
\begin{itemize}
\item 61 instructions
\item 3 double word loads (128 bit wide instructions (LLDW))
\item 4 branch instructions and single return
\item 4 loads of which 2 in the main loop.
\item the rest is arithmetic
\end{itemize}
\fi

\subsection{Networked Benchmark Setup}

Next, we construct a setup involving a simple IoT networked application as case study.
The \ac{VM} hosts high-level logic, and this loaded application is updateable over the network.
The functionality mimics that studied in prior work~\cite{riot-js-container} using small JavaScript run-time containers hosting application code on top of RIOT\@.
The hosted logic has access to both the CoAP stack and the high-level sensor interface (SAUL) provided by  RIOT~\cite{baccelli2018riot}.
The VM execution is triggered by a CoAP request and the operating system expects a formatted CoAP response payload or an error code from the application loaded in the \ac{VM}.
The goal is to load an application into the \ac{VM} that, when triggered by a CoAP request, reads a sensor value and constructs a full CoAP payload as response to the requester.

\section{Experimental Measurements}
\if{0}
Interpreter memory requirements explained:
\paragraph{rBPF}: 512B for stack + 88 byte for the registers + 60 byte context,
              and 16B per memory region.
\paragraph{WASM3}: 85KiB (64KiB for the stack, 11 KB for the compiled application, 4096 for stack, a number of misc other structures)
\fi

Using our experimental setup, we carried out an initial set of measurements comparing rBPF and WASM3\@.
With each prototype, we measured the performance of VM logic when it hosts the same Fletcher32 checksum.
While this example is specific and artificial, it is a good guinea pig to get an idea of what to expect in general.
The Fletcher32 checksum algorithm requires a mix of mathematical operations memory reads and branches, containing a loop over input data.

First and foremost as visible in \autoref{tbl:vmsize}, we observe that the Flash memory footprint of the interpreter WASM3 is 15 times bigger than the rBPF interpreter.
To get a perspective: relatively to the whole firmware image (assuming simple business logic and a CoAP/UDP/6LoWPAN network stack)  adding an rBPF \ac{VM} represents negligible Flash memory overhead (less than 10\% increase), whereas adding a Wasm \ac{VM} more than doubles the size of the firmware image.

\begin{table}[ht]
  \centering
  \begin{tabular}{lrr}
    \toprule
    & ROM size & RAM size \\
    \midrule
    WASM3 Interpreter  & \SI{64}{\kibi\byte} & \SI{85}{\kibi\byte} \\
    rBPF Interpreter   & \SI{4364}{\byte} & \SI{660}{\byte} \\
    Host OS Firmware (without VM)   & \SI{52760}{\byte} & \SI{14856}{\byte} \\
    \bottomrule
  \end{tabular}
  \vspace{0.5em}
\caption{Memory requirements for WASM3 and rBPF interpreters.}
\label{tbl:vmsize}
  \vspace{-1.5em}
\end{table}

Second, we give an initial measure of the data that needs transfer over the network when modular software update is performed (when \ac{VM} logic is updated).
With the results as in \autoref{tbl:scriptsize},
we observe that Wasm script size seem somewhat smaller than rBPF script size (approximately 30\% less in this case). 
The native C compilation shows the size of the code if the library is compiled into the device firmware itself and is not network updateable.

\begin{table}[ht]
  \centering
  \begin{tabular}{lrr}
    \toprule
    & code size & time \\
    \midrule
    Native C & \SI{74}{\byte} & \SI{27}{\micro\second} \\
    WASM3  & \SI{322}{\byte} & \SI{980}{\micro\second} \\
    rBPF   & \SI{456}{\byte} & \SI{1923}{\micro\second} \\
    \bottomrule
  \end{tabular}
  \vspace{0.5em}
  \caption{Size and performance of different targets for the fletcher32 algorithm}
  \label{tbl:scriptsize}
  \vspace{-1.5em}
\end{table}

Third, we compare the penalty in terms of execution time for VM logic. 
We measured the performance of Fletcher32 computation on a sample input string of \SI{361}{\byte}, with each VM\@.
We observe that execution is longer with the rBPF \ac{VM}, than with the Wasm \ac{VM} (2 times longer).
Both \acp{VM} perform significantly slower than native execution, with WASM3 approximately 35 times slower and rBPF around 70 times slower.
While this relative overhead is significant, the absolute overhead is not significant for hosted logic that is not computation-intensive.
Furthemore, in terms of instructions, rBPF still enables 1.3M instructions per seconds -- enough for a low-power IoT device, which is generally not required to process ultra-high data throughput.

Based on these preliminary measurements, we can conclude that rBPF seems to offer acceptable performance in general, and in particular a very substantial advantage in terms of Flash memory footprint compared to Wasm\@.
Hence, a VM approach based on rBPF seemed promising, and we 
have thus fleshed out our prototype further, to perform additional experiments with IoT use-cases involving a CoAP network stack, which next we report on.

\subsection{rBPF with Logic involving IoT Networking}

We here reproduce a use-case described in prior work~\cite{riot-js-container},
whereby high-level logic involving CoAP networking is executed by the \ac{VM}.
More precisely, we evaluated the performance the hosted code shown in \autoref{lst:coap}.
The application requests a measurement value from the first sensor and stores the value in a CoAP response.
The functions called from the application are provided as bindings by the host operating system and exposed to the \ac{VM}.
We implemented the CoAP bindings and as well as the bindings to the high-level sensor interface (SAUL) as depicted in \autoref{fig:integration}.

\noindent\begin{minipage}[t]{\columnwidth}
\begin{lstlisting}[language=C,caption={Example networked sensor read application},label={lst:coap}]
int coap_resp(bpf_coap_ctx_t *gcoap)
{
    /* Find first sensor */
    bpf_saul_reg_t *sens = bpf_saul_reg_find_nth(1);
    phydat_t m; /* measurement value */

    if (!sens || 
            (bpf_saul_reg_read(sens, &m) < 0)) {
        return ERROR_COAP_INTERNAL_SERVER;
    }

    /* Format the CoAP Packet */
    bpf_gcoap_resp_init(gcoap, COAP_CODE_CONTENT);
    bpf_coap_add_format(gcoap, 0);
    ssize_t pdu_len = bpf_coap_opt_finish(gcoap,
            COAP_OPT_FINISH_PAYLOAD);

    /* Add the sensor as payload */
    uint8_t *payload = bpf_coap_get_pdu(gcoap);
    pdu_len += bpf_fmt_s16_dfp(payload, m.val[0],
                               m.scale);
    return pdu_len;
}
\end{lstlisting}
\end{minipage}

When compiled, the size of the bytecode is \SI{296}{\byte}.
The overhead of the full script execution, including the execution of the function calls, is \SI{94}{\micro\second}.
The additional overhead caused by the \ac{VM} is negligible, when compared to network latencies of several milliseconds.

The size of the full firmware image is \SI{69}{\kibi\byte}, including the rBPF interpreter.
While the Flash memory required for the core rBPF interpreter is identical to the previous example (see Table \ref{tbl:vmsize}), there is however an \SI{80}{\byte} increase in Flash size due to the additional bindings to the CoAP and sensor interfaces.
The RAM requirements are increased by \SI{16}{\byte} for an additional memory access region, used to allow access to the CoAP packet.

Here, as an additional point of comparison, we can refer to similar logic hosted in a small embedded JavaScript run-time container with RIOT bindings, studied and measured in~\cite{riot-js-container} on similar hardware (a Arm Cortex-M microcontroller).
These measurements show that similar logic requires \SI{156}{\kibi\byte} for the JavaScript engine, on top of the \SI{59}{\kibi\byte} used by RIOT, and the hosted code (script) size which was around \SI{1}{\kibi\byte}. Note furthermore that these JavaScript containers did not specific memory isolation guarantees, as does rBPF\@.
We can thus conclude that rBPF offers much better prospects than embedded JavaScript run-time containers too, in terms of memory requirements, hosted logic size (and network traffic overhead required to transmit VM updates).

\section{Discussion \& Next Steps}




\paragraph{Inherent Limitations with a VM}
By construction, a \ac{VM} causes execution overhead, and thus increased power consumption, for logic executed within the \ac{VM}.
Measuring the full impact of the \ac{VM} on power consumption is a complex task.
However, this impact is mitigated by two factors.
On one hand, depending on the characteristics of the logic executed in the \ac{VM}, this overhead may be negligible.
Here, we gear the \ac{VM} towards hosting simple scripts, implementing short decision steps rather than lengthy bulk data processing.
In such cases, the additional power consumed is not substantial.
On the other hand, smaller script size decreases drastically the energy needed otherwise to transfer software updates -- 
especially compared to an alternative such as firmware updates, as shown in~\cite{riot-js-container}. 


\paragraph{Decreasing Wasm RAM usage}
WebAssembly has large RAM requirements: \SI{64}{\kibi\byte} memory pages increment is big for microcontrollers.
The WASM3 interpreter also adds an intermediate compile step, which increases speed, and collaterally RAM usage, by another \SI{10}{\kibi\byte}.
We thus cannot conclude just yet on how useful Wasm really is for low-power IoT.
An adaptation skipping this step and/or using smaller memory pages increments could reduce RAM requirements.
Next steps here could also include trying out other Wasm interpreters, such as for instance Wasm-micro-runtime~\cite{wasm-micro-runtime} and WARDuino~\cite{gurdeep2019warduino}.

\paragraph{Improving rBPF execution time overhead}
If execution time overhead is an issue, an option is to design from scratch a solution going beyond software-only, using hardware \ac{MPU} or even an MMU as base.
Another option is adding an intermediate transpilation technique to rBPF (similar to what is used by WASM3)
translating the raw eBPF instructions to a format more suitable for direct consumption on the system.
A more advanced step would be to translate these, ahead-of-time, into native instructions on the embedded device.

\paragraph{Decreasing rBPF script size overhead}
The rBPF \ac{VM} implementation is designed as a secure sandbox for running untrusted code on small embedded devices while adhering to the already defined eBPF ISA\@.
It can be seen from the application script sizes that the current implementation are relative big compared to applications compiled to WebAssembly.
As the eBPF instructions are fixed in size and can contain a lot of unused bit fields depending on the exact instruction, compressing them with well known algorithms can solve this downside.
Initial measurements show that Heatshrink~\cite{heatshrink}, an LZSS-based~\cite{lzss_compression} compression library suitable for small embedded systems, can reduce the application size by \SI{60}{\percent} depending on the application surpassing similar WebAssembly applications.

\paragraph{Extending rBPF sandboxing guarantees}
The current use case of rBPF targets execution of small-sized business logic and debug applications.
However the current VM design does not limit the actual execution time of the application: a virtualised application can keep the system busy without limitations, possibly draining the battery of the IoT device.
A potential next step could be to cap the CPU time which a single invocation of the virtual machine can occupy.

\if{0}
As shown the rBPF interpreter can be embedded into a system without significantly increasing the size of firmware.
Execution speed however suffers from the overhead caused by the interpreter.
An intermediate transpilation technique similar to what is used by WASM3 can be added,
translating the raw eBPF instructions to a format more suitable for direct consumption on the system.
A more advanced step is to translate them ahead-of-time into native instruction on the embedded device.
Either of these enhancements however will increase the memory requirements.

The rBPF \ac{VM} implementation is designed as a secure sandbox for running untrusted code on small embedded devices while adhering to the already defined eBPF ISA\@.
It can be seen from the application script sizes that the current implementation are relative big compared to applications compiled to WebAssembly.
As the eBPF instructions are fixed in size and can contain a lot of unused bit fields depending on the exact instruction, compressing them with well known algorithms can solve this downside.
Initial measurements show that Heatshrink~\cite{heatshrink}, an LZSS-based~\cite{lzss_compression} compression library suitable for small embedded systems, can reduce the application size by \SI{60}{\percent} depending on the application surpassing similar WebAssembly applications.

One of the limitations of the chosen WebAssembly here implementation is the extensive RAM requirements.
The cause here is twofold, first there is the WebAssembly specification dictating \SI{64}{\kibi\byte} memory pages.
The second reason is the intermediate compile step used by the WASM3 implementation increased the RAM usage by \SI{10}{\kibi\byte},
excluding this step in an interpreter can trade a reduction in execution speed performance for reduced memory consumption.
An implementation more geared towards embedded applications might be able to reduce the RAM requirements.

A different direction is to use hardware capabilities of different platforms for sandboxing.
Instead of using an interpreter using eBPF or WebAssembly code, native code can be sandboxed by leveraging mechanisms such as an \ac{MPU}.
While similar care has to be taken to ensure that the sandbox is secure, the performance of the system can be much higher because of the lack of an interpreter.
\if{0}
\begin{itemize}
\item WASM3 is not the only interpreter, wasm-micro-runtime might be more suitable for embedded
\item Wasm is unsuitable? for microcontrollers due to the 64KiB heap chunks.
\item MPU for more efficient native sandboxing.
\end{itemize}
\fi
\fi

\section{Conclusion}

In this paper we have designed, implemented and studied experimentally two minimal \aclp{VM} targeting low-power, microcontroller-based IoT devices.
We designed rBPF\@, a register-based VM hosted in RIOT, and an interpreter, based on Linux's extended Berkeley Packet Filters.
We compared its performance, experimentally on commercial IoT hardware, to an approach hosting and isolating logic in an embedded WebAssembly virtual machine.
We show that, compared to WebAssembly and to prior work on small run-time containers for interpreted logic,
rBPF is a promising approach to host and isolate small software modules,
yielding acceptable overhead in execution time,
and very small memory overhead (approx. 10\%) for a typical IoT application.


\ifCLASSOPTIONcaptionsoff
  \newpage
\fi

\bibliographystyle{abbrv}
\bibliography{bibliography}

\balance




\end{document}